# QoS-based Packet Scheduling Algorithms for Heterogeneous LTE-Advanced Networks: Concepts and a Literature Survey

Najem N. Sirhan, Manel Martinez-Ramon

Electrical and Computer Engineering Department,
University of New Mexico, Albuquerque, New Mexico, USA

## ABSTRACT

*The number of LTE (Long-Term Evolution) users and their applications has increased significantly in the last decade, which increased the demand on the mobile network. LTE-Advanced (LTE-A) comes with many features that can support this increasing demand. LTE-A supports Heterogeneous Networks (HetNets) deployment, in which it consists of a mix of macro-cells, remote radio heads, and low power nodes such as Pico-cells, and Femto-cells. Embedding this mix of base-stations in a macro-cellular network allows for achieving significant gains in coverage, throughput and system capacity compared to the use of macro-cells only. These base-stations can operate on the same wireless channel as the macro-cellular network, which will provide higher spatial reuse via cell splitting. Also, it allows network operators to support higher data traffic by offloading it to smaller cells, such as Femto-cells. Hence, it enables network operators to provide their growing number of users with the required Quality of Service (QoS) that meets with their service demands. In-order for the network operators to make the best out of the heterogeneous LTE-A network, they need to use QoS-based packet scheduling algorithms that can efficiently manage the spectrum resources in the HetNets deployment. In this paper, we survey Quality of Service (QoS) based packet scheduling algorithms that were proposed in the literature for the use of packet scheduling in Heterogeneous LTE-A Networks. We start by explaining the concepts of QoS in LTE, heterogeneous LTE-A networks, and how traffic is classified within a packet scheduling architecture for heterogeneous LTE-A networks. Then, by summarising the proposed QoS-based packet scheduling algorithms in the literature for Heterogeneous LTE-A Networks, and for Femtocells LTE-A Networks. And finally, we provide some concluding remarks in the last section.*

## KEYWORDS

*HetNets LTE-Advanced networks, Packet scheduling algorithms, QoS.*

## 1. INTRODUCTION

### 1.1. Quality of Service (QoS) in LTE

QoS refers to the ability of delivering a service with a required quality level that meets with the customer expectations, such as making a video call without distortion or disconnecting. LTE delivers a variety of data types using limited air resources and routing interfaces of the network. Since LTE is an IP based network, there are no dedicated switch circuits that are assigned to an active session and therefore an alternative approach has to be followed or applied in order to guarantee the required quality of service. This approach is based on attaching a QoS tags

"parameters" to each packet, these tags allow the network to differentiate between the customers and also among the services [23].

There are three main principles for applying QoS in an LTE cellular network. The first principle is to differentiate between services, and between subscribers. The second principle is prioritization, which depends on the differentiation, it aims to provide a required priority to each customer while allocating air interfaces and routing resources at each node of the network. The third principle is admission control, it aims at controlling which service is going to be served or not, and this is based on its priority, for example, services with lower priority might be blocked in the case of scarce resources [23].

There are two to four QoS parameters for a bearer, it depends on whether the service is real time or best effort service, and they are [11]:

- Allocation and Retention Priority (ARP): this tag is used as a mechanism to drop or downgrade lower-priority bearers in network congestion scenarios. It is also used in bearer establishment, and has a high significance in handover scenarios in which the networks checks it when determining if new dedicated bearers can be established through the eNodeB.
- Quality of Class Indicator (QCI): this tag determines the treatment of IP packets received on a specific bearer. The value of QCI affects several node related parameters, such as link layer configuration, scheduling, and queue management. 3GPP standardised QCI attributes are shown in Table 1.
- Guaranteed Bit Rate (GBR) for real time services only.
- Maximum Bit Rate (MBR) for real time services only.

Table 1. 3GPP standardised QCI attributes [11].

| QCI | Resource Type | Priority | Packet Delay Budget | Packet Error Loss Rate | Example Services |
|-----|---------------|----------|---------------------|------------------------|------------------|
| 1 | GBR | 2 | 100ms | $10^{-2}$ | Conversational voice |
| 2 | | 4 | 150ms | $10^{-3}$ | Conversational video (live streaming) |
| 3 | | 3 | 50ms | $10^{-3}$ | Real-time gaming |
| 4 | | 5 | 300ms | $10^{-5}$ | Non-conversation video (buffered streaming) |
| 5 | Non-GBR | 1 | 100ms | $10^{-3}$ | IMS signaling |
| 6 | | 6 | 300ms | $10^{-5}$ | Video (buffered streaming) TCP-based (email, chat, file sharing .. etc.) |
| 7 | | 7 | 100ms | $10^{-5}$ | Voice, video (live streaming), interactive gaming |
| 8 | | 8 | 300ms | $10^{-3}$ | Video (buffered streaming) |
| 9 | | 9 | 300ms | $10^{-5}$ | TCP-based (email, chat, file sharing .. etc.) |

In Table 1, Priority parameter provides the scheduling and routing priority of each packet at each node. Also, Delay parameter is the upper bound of the delay time that could be experienced between the User Equipment (UE) and the Packet Gateway (P-GW). Also, Packet Error Loss Rate (PELR) parameter is the the upper bound of the percentage of the packets that might be lost.

In order to manage the delivery of packets in a differentiated manner, packets are grouped into bearers. A Bearer is defined by the combination of QoS class and a destination IP address. A bearer may include packets belonging to different services, as long they require the same QCI and the same UE. There are two main types of bearers, Guaranteed Bit Rate (GBR) and Non-Guaranteed Bit Rate (N-GBR). In the case of *GBR bearers*, they are established on demand because a minimum amount of bandwidth is reserved by the network regardless of whether it is used or not during the admission control function, inactivity timers are used to control air interface to free up resources. The precedence of service blocking over service dropping in congestion scenarios, GBR bearer should not experience any packet loss on the IP network or the radio link due to congestion. GBR bearers are used for real time services. In the case of *N-GBR bearers*, there are no specific reserved bandwidth so they can remain established for long periods of time, Precedence of service dropping over service blocking in congestion scenarios, they are treated with lower priority than the N-GBR bearers, they experience packet loss during congestion, they are used for best effort services [11][2]. Uplink bearers are created at the UE. While the creation of downlink bearers depends on the type of used protocol, whether it is GPRS Tunnelling Protocol (GTP) or Proxy Mobile IPv6 (PMIP). In the case of GTP, the downlink bearers are created at the P-GW, more specifically at Policy and Charging Control Function (PCRF). In the case of using the PIMP, the downlink bearers are created at the S-GW more specifically at the Binding and Error Reporting Function (BBERF) [23].

The PCRF is the policy server in the Evolved Packet Core (EPC). The PCRF takes the available network information and operator-configured policies to create service session-level policy decisions. The decisions, known as Policy and Charging Control (PCC) rules "PCC block diagram is shown in Figure 1", are forwarded to the Policy and Charging Enforcement Function (PCEF) located in the Packet Data Network Gateway (PDN-GW). Part of the PCC rules is the Traffic Flow Templates (TFTs) as shown in Figure 2. The PCEF enforces policy decisions by establishing bearers, determines which packet flows are mapped into each dedicated bearers, and performing traffic policing and shaping as shown in Figure 3 [11][2].

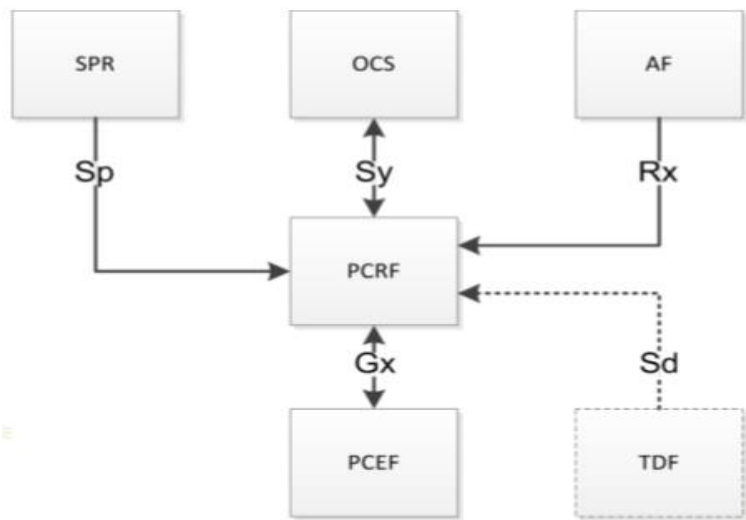

Figure 1. PCC block diagram [17]

The six main functions of the PCC are shown in Figure 1 and they are; Subscription Profile Repository (SPR), Online Charging System (OCS), Application Function (AF), Policy Charging Rules Function (PCRF), Policy Charging Enforcement Function (PCEF), Traffic Detection Function (TDF) [17].

The PCEF is the main component of PCC, and its use is mandatory. An operator can have pre-provisioned PCC rules in the PCEF. The PCEF act as a gateway for services, it allows a service data flow, that is subject to policy control, this provides a means of blocking unknown or unenforced traffic. It acts as a Charging Trigger Function (CTF) where through Diameter Credit Control (DCC) it feeds information to an OCS in order to track usage. It also act as a Charging Data Function (CDF) through offline charging records required for typical post-paid services and charging reconciliation. And it also enforces QoS, it converts a QoS class identifier value to IP-session specific QoS attribute values and determine the QoS class identifier value from a set of IP-session specific QoS attribute values. It also enforces the authorized QoS of a service data flow according to the active PCC rule [17].

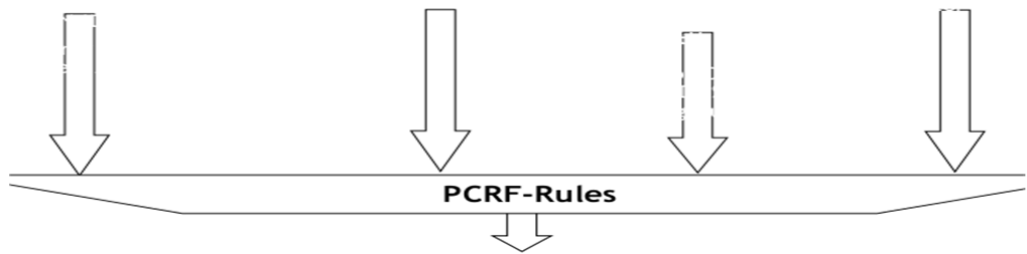

Figure 2. PCRF output forms the PCC rules [2]

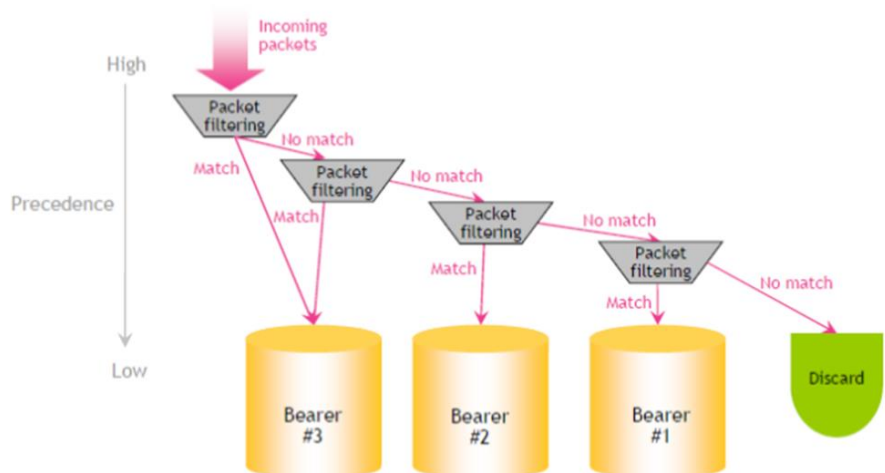

Figure 3. Traffic Flow Templates [2]

## 1.2. Heterogeneous LTE-A Networks (LTE-A Hetnets)

Heterogeneous Networks (HetNets) include a mix of macro-cells, remote radio heads, and low power nodes such as pico-cells, and femto-cells [19].

Macro-cells are high power eNodeB with a coverage of few kilometres, it provides an open public access with guaranteed minimum data rate under a maximum tolerable delay, it uses a dedicated back-haul that is capable of emitting power up to 46 dBm [18].

Remote Radio Head (RRH) are compact-size, high-power and low-weight units, which are mounted outside the conventional macro-cell's base station, and connected to it through a fibre optic cable to create a distributed base station, in-which the central macro-cell's base station controls the baseband signal processing, and moves some of the radio circuitry into the remote antenna. The use of RRHs reduces the power consumption by eliminating the power losses in the antenna cable [21].

Pico-cells are low power eNodeBs with a coverage of 300 meters, they are usually deployed in a centralized way with the same back-haul and access features as macro-cells, they are deployed in outdoor or indoor coverage, and they are capable of emitting power between 23 to 30 dBm [20]. Femto-cells are also called as home base stations, they are indoor base stations that are installed in homes and offices for getting better coverage and capacity gain. They provide better coverage due to the short distance between the transmitter and the receiver "about 50 meters at max" which reduces the power consumption. They provide better capacity gain by achieving higher Signal to Interference plus Noise Ratio (SINR) through the use of dedicated base stations to its users [19] [20].

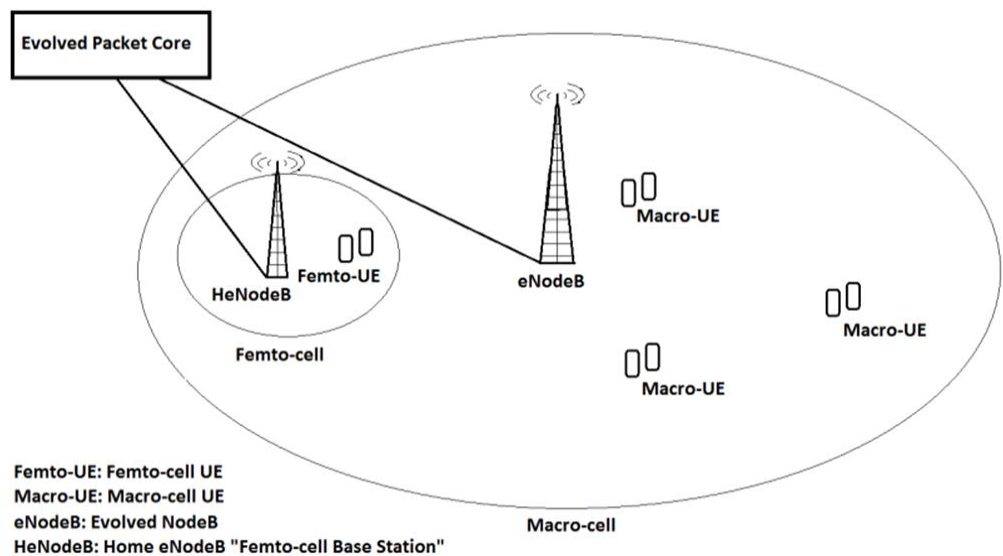

Fig 4. A basic model that represent the LTE-A HetNets which consists of a macro-cell and a femto-cell, and how they are connected to the LTE core network [21]

## 2. TRAFFIC CLASSIFICATION FOR QOS LTE-A PACKET SCHEDULING ALGORITHM

In [12], they propose two Service Specific Queue Sorting Algorithms (SSSA) one for Real Time (RT) and the other for Non Real Time (NRT) streaming video traffic. They implemented their SSSA in a QoS aware dynamic Packet Scheduling Architecture (PSA) that is designed by [13], which is shown in Figure 5.

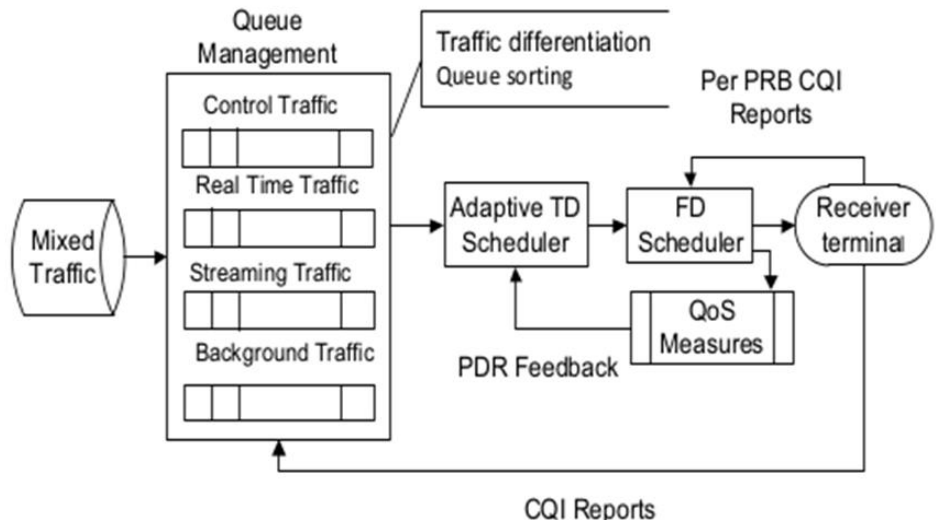

Fig 5. Packet scheduling Architecture designed by [13]

In [13], the packet scheduling architecture that they designed is suitable to be used for real-time traffic due to its ability to differentiate mix traffic into service specific queues, and it also sort users into queues, and adaptively reserves available resources to real time and non-real time traffic types. This scheduler adds two functionalities to the LTE-A downlink transmission. The first function is service specific queue sorting algorithms for service level performance optimization. The second function is adaptive Time Domain (TD) prioritizing algorithm for network level optimization.

In [13], for the case of real time traffic this approach significantly reduces delay and packet Drop Rate (PDR), while satisfying minimum throughput requirements of non-real time streaming video traffic. This is done by slightly delaying Best Effort (BE) traffic. In the case of real time video streaming, a ratio of the instantaneous and average achieved throughput is taken into account to prioritize users with lower average achieved throughput. Also the normalized waiting time is taken into account.

The SSSA that was designed by [12] consists of Traffic differentiator, TD scheduler, and FD scheduler. The traffic differentiator classifies the incoming traffic into four classes and sorts users according to their traffic type. The four traffic classifications are, control traffic "e.g. control information", real time traffic "e.g. voice calls", non real time traffic "e.g. streaming traffic", and background traffic "e.g. Best Effort traffic such as e-mail ". The TD scheduler picks a pool of users from these queues and then the FD scheduler allocates Physical Resource Blocks (PRB) to the users. These queues represent the QoS requirements for each traffic type. In the case of control traffic queue, the control traffic is equally important for all users, therefore users are sorted in round robin fashion. In the case of BE traffic queue, the BE traffic does not have a QoS requirements.  In the case of real time traffic queue, the delay of real time traffic has to be maintained less than a delay budget, which is defined to as an upper bound of delay for real time traffic. The RT users are being sorted by an algorithm that is the product of normalized waiting time of each user and its channel conditions, then the product is added into the queue length of the user.  Packets in the queue are being normalized by the queue algorithm by means of arranging them according to their delays, the packet with the longest delay comes first. Normalizing the packets has the benefit of prioritizing them to reduce the Packet Drop Rate (PDR) due to time out. As a result the delay will be reduced, fairness among users will be improved, and the system's overall throughput will be improved. In the case of non real time traffic queue, the QoS requirement for non real time streaming video traffic is defined based on making the instantaneous throughput of a user as an upper bound to its non real time throughput requirement. The metric that was used is the product of normalized waiting time, which is a ratio of minimum required throughput and average achieved throughput, and channel conditions of each user. The use of normalized waiting time will result in improving the fairness among users by equalizing their waiting time and reducing their delay.

## 3. QOS-BASED PACKET SCHEDULING FOR LTE-A HETNETS

The deployment of hybrid radio access has two main challenges: providing effective QoS and fair admission control. Motivated by these challenges, [4] propose a traffic-aware OFDMA hybrid small-cell deployment for QoS provisioning and an optimal admission control strategy for next-generation cellular systems. In order to do so, the traffic awareness in their proposed framework is accomplished by deriving a novel traffic-aware utility function, that differentiates the user QoS levels with the user's priority indexes, channel conditions, and traffic characteristics. They also propose an admission control algorithm based on their utility function. They tested their proposed framework, and their results showed that it achieved an optimum QoS performance in terms of total throughput and traffic delay.

In [4], their proposed framework "as shown in Figure 6" consists of four main parts and is embedded into each Small-cell Access Point (SAP). The four parts are, the QoS Classification of Heterogeneous Traffic, the Calculation of Utility Function, Traffic-Aware Admission Control, and Power Constraint Scheduling. In the QoS Classification of Heterogeneous Traffic part, the requests of the small-cell users (SUs) and the External Users (EUs) are used in order to calculate the average waiting time for each user type, these calculations results in classifying the traffic into different QoS types, such as the Best Effort (BE) traffic, video streaming traffic and the Constant Bit Rate (CBR) traffic. On the Calculation of Utility Function part, they define how the sub-carrier $k$ is allocated to user n, they have three different definitions, each one of them depends on the traffic QoS type. In the Traffic-Aware Admission Control part, they implemented an admission control algorithm that focus on balancing the load on the system that are based on a set of rules which are specific for performing admission control in a hybrid small cell with heterogeneous user traffic. In order for these rules to be applied, they defined certain priorities, one for the user type and the other for the traffic type, for example, for the user priority, the SUs has a higher priority over the EUs, and for the traffic priority, the priority decreases from CBR to video conferencing to BE. In the power constraint scheduling, they proposed a utility function with an optimized objective to allocate sub-carriers to users with power constraints.

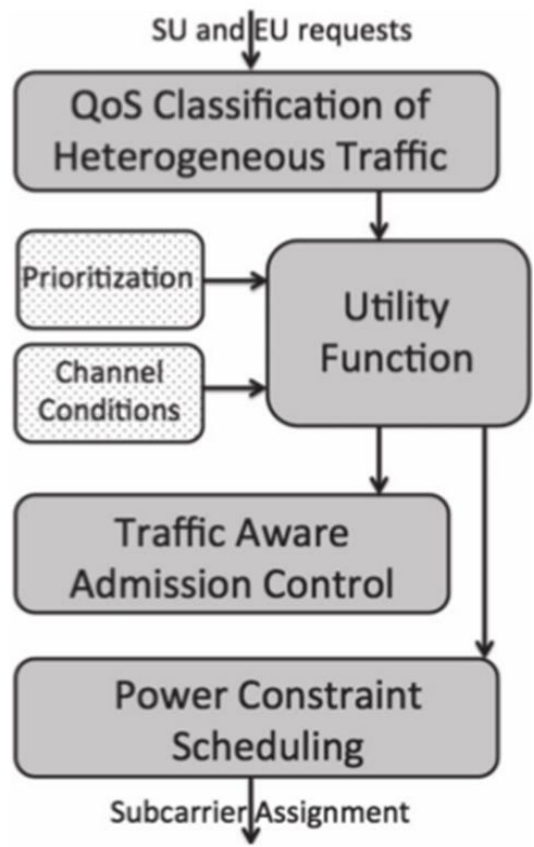

Fig 6. Framework proposed by [4]

In [4], their utility function was modelled based on OFDMA system parameters and queue models using MATLAB. Their minimal algorithm routine that was implemented in C was used through the MATLAB Mex file to perform sub-carrier allocation decisions. They compared three different schemes in terms of mean delay and delay variance of the worst case user of each traffic class, and they displayed their simulation results in plots. In Scheme 1, only the TA-Utility scheme was implemented, while In Scheme 2, the TA-Utility scheme was enhanced using the proposed admission control procedure. In the third scheme, the Maximum-Largest Weighted Delay First (M-LWDF) scheduling was used.

In [4], their simulation results for the case of CBR traffic and in terms of the mean delay for the worst case CBR user, for both Scheme 1 and Scheme 2 remained within 50 ms for arrivals below 300 KB/s when compared with a mean delay of 80 ms under the M-LWDF scheme. For arrivals beyond 300 KB/s, the M-LWDF scheme has the mean delay increasing almost linearly, while Scheme 1 and Scheme 2 resulted in a slowly increasing mean delay with Scheme 2 resulting in up to 50 ms less delay than Scheme 1 at arrival rates beyond 650 kB/s. The delay variance of CBR users for different arrival rates. In the case of the delay variance for their proposed schemes when compared with the M-LWDF scheme indicated that a higher degree of fairness is achieved for different CBR users.

In [4], their simulation results for the case of video traffic and in terms of the mean delay for the worst case video users, scheme 1 has a comparable mean queuing delay to M-LWDF for all arrivals below 350 KB/s. Beyond 350 KB/s, the mean delay was increasing slower than the M-LWDF scheme. Scheme 2 resulted in further improvement from Scheme 1 with the maximum mean delay reaching up to 190 ms. Therefore, the TA-Utility scheduling offers significant delay performance gains for delay-sensitive traffic classes. In the case of the delay variance, it was lower for both Scheme 1 and Scheme 2 when compared with that of the M-LWDF scheme, showing good fairness performance.

In [4], their simulation results for the case of BE traffic and in terms of the mean delay for the worst case BE user, the mean delay for BE users was much higher and reached up to 1s for both Scheme 1 and Scheme 2 when the arrival rate was 600 KB/s and above. In the case of the delay variance, it was significantly higher for low data rates for their proposed schemes, particularly for Scheme 1. However, as the arrival rates of BE users increased, the delay variance converged toward that of the M-LWDF scheme.

In [3] the authors propose and evaluate an optimal scheduling scheme for QoS provisioning for hybrid small cells. This framework takes into account the power constraint in addition to the user's priority index and traffic characteristics in order to efficiently provide differentiated QoS benefits to users served under an OFDMA hybrid small cell.

In [3], the authors briefly introduced the difference between the three small cell access polices, the open access, closed access, and hybrid access policy. The authors recommended the use of the hyper access policy which is a mix of the first two polices, and it provides a differentiated service to the Small Cell Users (SUs) and Macrocell Users (MUs). This is because it provides a guaranteed QoS as in the closed access policy, and the capacity enhancements as in the open access. They also explained how a previous two hybrid policy failed to consider the nature of the higher layer traffic in performing scheduling and access control for small cells.

In [3], the authors explained their system model for the OFDMA based hybrid small cells. They started by presenting and a briefly reviewing some of the current scheduling and whether they are suitable to be used as a scheduling mechanism in the small cells or not. For example, the PF and the M-LWDF are not suitable because they don't provide bounded delay performance.

In [3], the authors presented the formula of their utility function that is associated with the allocation of subcarrier k to user n, and how the formula parameter values could be changed to represent different user and traffic types. Then they explained their optimization objective of subcarrier allocation with power constraints using their proposed utility function. Then they solved their optimization objective after classifying it into the Multiple Choice Knapsack Problem (MCKP).

In [3], the authors explained how they evaluated the performance of their system using MAT-LAB and the minimal algorithm routine is implemented in C. They classified users based on their traffic into three classes, the Constant Bit Rate (CBR) users, Video Streaming users, and the Best Effort (BE) users. Then they explained their simulation scenario and its parameters which they used in order to measure the throughput performance and the delay. They measured the time average throughput performance that was achieved by the users, and they represented it in bar graphs. They also presented the mean delay vs. arrival rate performance of their proposed solution for the three user types in comparison with the M-LWDF and they represented their results in line graphs. According to their graphs, it was clear that their utility could achieve lower and bounded delay for delay-sensitive traffic types such as the video streaming and CBR when compared to the M-LWDF. However, the BE traffic experienced large delays when compared to the M-LWDF but it remained in the acceptable range. Finally, the authors drew their conclusions and they stated that their proposed work could be used as a framework in order to design efficient admission control algorithms in a hybrid small cell network.

In [8], the authors introduced the concept of Cognitive Radio (CR) based small cells and how they cluster users that utilize different channels in order to prevent the occurrence of a communication bottleneck by the use of dynamic spectrum access techniques. They also introduced the two main challenges in the deployment of such cells. The first challenge is in detecting the available channels by performing an accurate spectrum sensing and monitoring. The second challenge is to maintain a reliable topology control which is directly related to the available spectrum and the utilization of the licensed channels.

In [8], the authors introduced the system architecture and their proposed framework. They explained their scenario's network architecture, communication nodes, and the traffic type. They used a figure that shows the effect of applying their topology in clustering the users. They explained the two main parts that their proposed framework which is located in the CR Base Station consists of, the Monitoring and the Assignment. The Monitoring module gathers the CR requests' information such as the spectrum utilization, packet losses, delay and jitter) and also the background traffic information. The Assignment module responsibilities relies on analysing the information gathered by the monitoring module in order to make a decision for the spectral topology assignment based on a formula that represent the relationship between traffic density and throughput utilization.

In [8], the authors explained how they created and evaluated the performance of their proposed framework by the use of ns-2. They also explained in detail the performance parameters that they used such as the latency, throughput, packet loss, and jitter and spectrum utilization. For each one of these parameters they studied the effect of incrementing the number of frequencies under different traffic densities. Their results showed that the system's performance improved in terms of all these parameters when the number of frequencies was increased.

The authors of [14] proposed a multi-hop wireless network that uses the Any-cast Back-pressure (AB) routing "a practical distributed any-cast routing protocol designed to scale with the number of gateways and to exploit path and gateway diversity" as a scalable mobile back-haul for dense small cell deployments. Also the authors explained how their proposed solution uses the any-cast back-pressure routing. Firstly, they explained the operation of distributed any-cast back-pressure routing and how it finds the appropriate trade off between getting as geographically close as possible to the destination and how it evenly distributes the load among all neighbours by exploiting queue backlog differential information. Secondly, they explained how they took into considerations the 3GPP data plane architecture. Finally, they explained the flexibility of the AB in the deployment of new gateways and with the addition of new capacity.

Also in [14], the authors explained how they used ns-3 to evaluate and compare the performance of their routing solution and two other unicast routing solutions, the Unicast-multipath and the Unicast-Shortest Path. Their comparative evaluation was based on a 5x5 grid of femtocells scenario, they considered two different back-haul settings in their simulations, homogeneous and heterogeneous transmission rates, and they measured the throughput and latency for both settings. In the case of homogeneous link rates scenario, they measured the throughput and delay when the number of gateways increased from one to five. Their simulation results showed that the AB outperformed the other uni-cast routing solutions. In the case of heterogeneous link rates scenario, they evaluated how the three routing protocols utilized the aggregated capacity offered by five gateways under wireless link dynamically in order to study the dependency of the aggregated throughput and latency on the percentage of links using the lowest 802.11a rate (i.e., 6Mbps). Their simulation results showed that their solution outperformed the other uni-cast routing solutions, so the authors defended their choice of using the AB as a transport routing for dense small cell deployments.

## 4. QOS-BASED PACKET SCHEDULING FOR LTE-A FEMTOCELLS

According to [16], Resource Blocks (RBs) can be shared among several Femtocell users simultaneously, however, they can't be shared among macro users. This is because the RBs are orthogonal to each other in the case of macro cells, this means that there is no need for the interference coordination among macro users. However, in the case where Femtocells are present, the RBs, which are being used by Femtocells are not orthogonal to the macro cell, so Interference coordination is needed among those users.

There are two coordination approaches for interference coordination, the inter-tier and intra-tier interference coordination. In the case of Inter-tier interference coordination strategy, the allocation of RBs between the macro-cell and Femtocell users is always orthogonal. The RB that is assigned to a macro user is not applicable to be reused. The RB that could be reused are the ones that are assigned to Femtocell users. In the case of Intra-tier interference coordination strategy, it is needed only for the Femto-tier where two floor models were considered, the Inter-floor and the Intra-floor models. In the case of Inter-floor modelling, a group of RBs is reserved for the fairness improvement of macro user, then the remaining RBs are equally divided into two groups. Each RB group is assigned to Femtocell users of the alternate floors. In the case of Intra-floor Modelling, Femtocell user can only reuse an RB that is served by another femtocell user when its Femtocell user's serving Femtocell BS is non-adjacent to the already assigned Femtocell user's serving Femtocell BS. The Femtocell BSs must be at least 10 meters apart in order for the RB to be reused, irrespective of Femtocell BS locations on the same floor, either in the same stripe or in different stripes [16].

The PF in a multi-user diversity is not efficient because it only allocates one RB for each user. In order to improve the PF to work in the addition of Femtocells, [16] has proposed a two-tier radio resource reuse and interference coordination technique that was employed to LTE-A Femtocells that uses the PF scheduling scheme as its baseline, they called it the Modified Proportional Fair (MPF). The idea of their scheduler is to facilitate the RB reuse among Femtocell users. Their scheduler doesn't only consider Femtocell users with the max PM but it also considers other Femtocell users for assigning RB in a TTI, subject to the interference coordination strategy. They applied their scheduler and tested its performance in an indoor environments. Their schedule is shown in Figure 7.

In the process of radio resource scheduling for macro and Femtocell users, usually Femtocell users have better coverage that result in a better signal quality. So in [16], they added the mechanism of the macro user RB reservation to provide fairness to macro users. It works by reserving a certain percentage of the total RBs of the system bandwidth for the macro users to be served in every TTI.

In [16], They compared their proposed MPF with and without the existence of interference to the regular PF in terms of average throughput and fairness. In terms of average throughput, their proposed MPF outperformed the regular PF in the absence of interference coordination. However, in the presence of interference it had a lower value average throughput. This is because an RB cannot be allocated blindly in the presence of interference. In terms of fairness, both the MPF and the PF had a similar performance. They also investigated the performance of the Macro user RB reservation in terms of the fairness of macro users as well as the average fairness performance of the system. Their simulation results showed that the average fairness performance of macro users and also the system performance was improved. They also investigated the performance of their proposed MPF scheduler in terms of spectral and energy efficiencies, varying the value of maxPM of MPF scheduler. Their simulation results showed that their proposed scheduler has the ability to meet the spectral efficiency requirements of LTE-Advanced systems taking into consideration the reduction in energy per bit assurance. So they suggest the use of their proposed resource reuse scheduler in the tier level in addition to the macro-cell level in order to improve both spectral and energy efficiencies that is able to meet with the highest data rate at low transmission energy per bit requirement of LTE-Advanced systems.

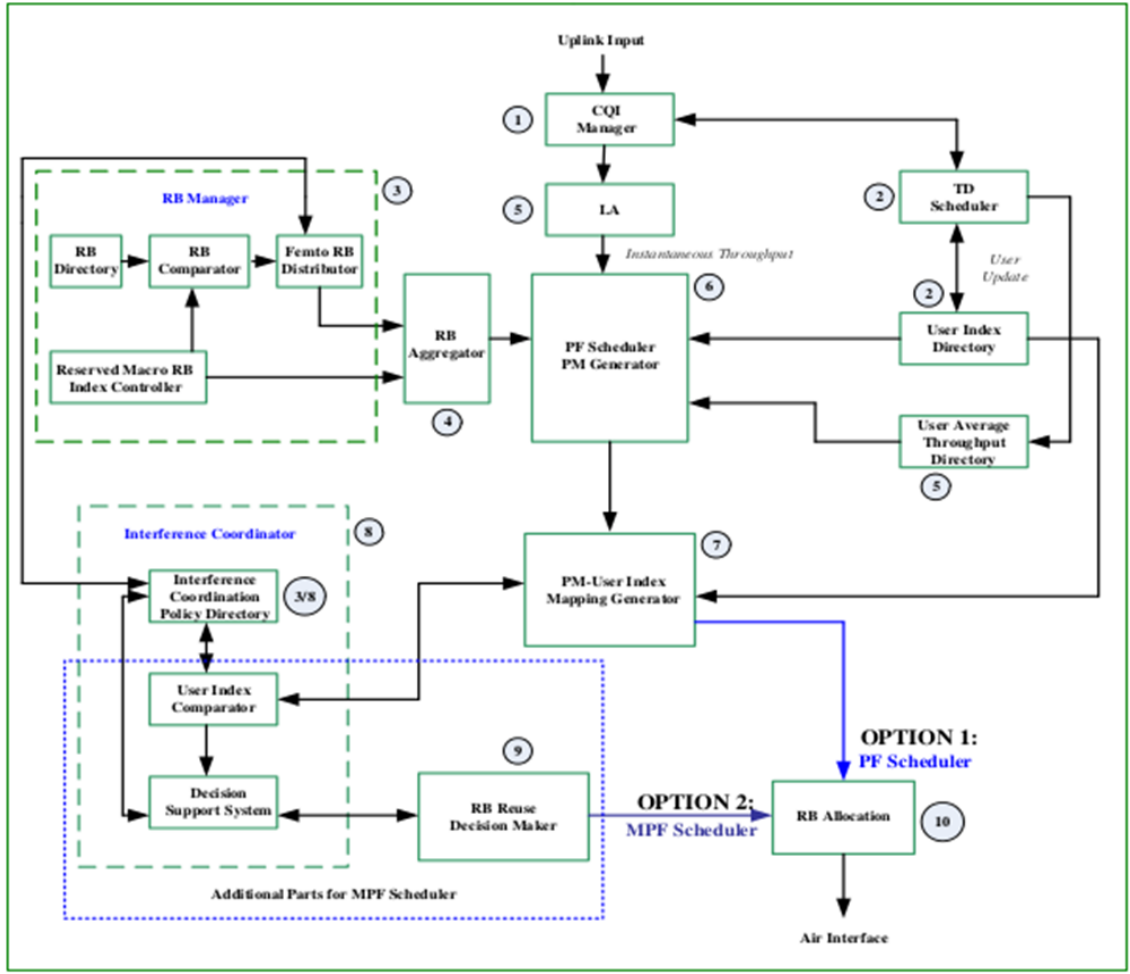

Fig 7. Resource scheduler proposed by [16]

The authors of [5] presented in their study a simulation tool for LTE femtocells, which was implemented as a module of the emerging open source LTE-SIM framework. It encompasses heterogeneous scenarios with both macro and femtocells, spectrum allocation techniques, user mobility, femtocell access policies, and several other features related to this promising technology. Also, the authors described the LTE, highlighted the pros and cons which are related to the development of this new technology and the most important open issues that justify the need of a simulation framework.

Also in [5], the authors described their developed module, with particular emphasis on the newly implemented propagation loss models for indoor scenarios, the introduction of new network topology objects, and the enhancements of some modules of the LTE-SIM framework. Also, they provided an overview about some of the possible studies that could be carried out using their simulation framework, such as; The impact that the co-layer interference due to the communication between home base station and user in an indoor scenario composed by a single 5×5 apartment grid. The impact of the femtocell deployment in urban environments. They designed a scenario consisting of one macro-cell and 56 buildings located as in a typical urban cross. They proposed a scalability analysis of their proposed module in terms of both simulation time and memory usage on a Linux machine with a 2.6 GHz CPU and 4 GBytes of RAM. These analysis were done by varying the number of UEs per building.

The authors of [6] presented and provided a detailed explanation about their module for simulating LTE femtocell within the LTE-Sim open-source framework, their module was designed to simulate Heterogeneous scenarios of macro-cells and Femtocells. They started their paper by introducing Femtocells and the need for their deployments to boost the network capacity, and to guarantee a better coverage for those areas which are currently provided with an unsatisfactory service if served by the macro-cells. Also, the authors described the framework by explaining what its main classes could handle. Its classes could handle new network devices, for example, handling Home eNodeB (HeNB) which could be configured for working with both open and closed access modes, and also an enhanced user equipments which could be recognized whether if they are inside or outside a building at certain times. Its classes could also handle handover procedures between macro-cells and Femtocells. Its classes could also handle new topology objects such as Femtocell, building and a street. And its classes could also handle new channel models in order to cope with the characteristics of indoor environment.

Also in [6], the authors displayed their numerical results in the form of bar figures. They studied how the aggregate cell throughput was affected by the frequency reuse factor under the probability that a single Femtocell is active. Their results showed that when a single Femtocell was active half the time, the aggregate cell throughput was best at frequency reuse 1/1, and when it was active all the time, the aggregate cell throughput was best at frequency reuse 1/2. They also studied how the network capacity was enhanced by the deployment of Femtocells. They deployed two different scenarios, the first scenario included buildings which are distributed in a macro-cell without Femtocells, and the second scenario is similar to the first, but in addition it had one Femtocells per apartment that was assumed to be active and working on the same operating bandwidth of the macro-cell. They compared the resulted throughput for both scenarios under different number of UEs per HeNB. Their results showed that not only the throughput was increased by the use of femtocells, but also it increased when the number of UEs was increased per HeNB.

The authors of [15] provided a quality based Call Admission Control and resource allocation mechanism to avoid resource overloading and call quality degradation. They investigated the problem of congestion, which could occur when a large number of Femtocells are using the Digital Subscriber Line (DSL) as a back-haul link. They briefly introduced the concepts of Self Organizing Networks (SONs), policy based network management, Femtocell networks and how voice calls are converted into Voice over Internet Protocol (VoIP) calls and encapsulated in a tunnel to be transmitted into the HeNBGW which is located in the operator's core network. Then, they briefly list the related work in this area, and how they differ from their own work in terms of how the voice traffic was emulated and also from where they located the measurement node which was responsible of measuring the voice call quality. Then, they described the network architecture of their deployed scenario that they used as a reference architecture in the development of their proposed admission control.

Also in [15], the authors described their proposed QoS based call admission control. Their description approach started by explaining how they measure the quality of on-going voice calls passing through the HeNBGW, they call this measurement value the Mean Opinion Square (MOS), this MOS value was mapped with a quality rating. The measurement of the MOS was based on calculating the packet delay, packet jitter, and packet loss. Then, they explained their Call Admission Control (CAC) algorithm by a flowchart that describes the decision making process. Then they explained how the DSL Access Multiplexer (DSLAM) monitors the bandwidth in the Expedited Forwarding (EF) queue and shapes the traffic by dropping any packets that arrives when the buffer is full. Then, they described how their proposed solution was implemented and validated by the use of ns-3 network simulator. They explained their deployed scenario and its parameters. At the end of their paper, the authors presented their results in figures, they studied how the MOS value was affected by the increase of the number of on-line calls versus time in two scenarios, in one scenario they used the CAC and dynamic resource allocation, and in another scenario they didn't use it. According to their results, in the scenario where their proposed solution was not used, the MOS dropped dramatically when the number of on-line calls increased. However, when their solution was used, the MOS maintained with a high quality rating.

In [1], the authors' main objective is to provide a unified multimedia-aware framework for downlink scheduling and radio resource management for the base station of an LTE Femtocell known as HeNB. This framework prioritize GBR traffic over NGBR traffic, and it aims to provide efficiency to GBR contents while at the same time preserving fairness for NGBR contents as well. The framework's algorithm is basically adaptive and opportunistic, for each traffic class, it assigns the Resource Blocks (RBs) to the users based on their Channel Quality Indicator (CQI) value. Then, they described their proposed model and how its algorithm works and acts as a MAC layer which enhances the efficiency of resource management process to be dynamic and opportunistic, and they described this by the use of workflow and a diagram. Then, they described how they implemented their framework and evaluated it by the use of ns-3 network simulator in particular using the LTE module developed by the LENA project.

The authors of [9], presented an architectural solution "it is transparent to the 3GPP architecture" for efficiently deploying Femtocells in the form of Networks of Femtocells (NoFs). The NoFs is designed in the framework of the Broadband Evolved FEMTO Networks (BeFEMTO) System architecture, it is a group of Femtocells in the same administrative domain that cooperates together for a global performance improvement. The key to this improvement is the introduction of a new entity called a Local Femto Gateway (LFGW) which acts as a proxy for Femtocells inside the NoF when establishing communications with the EPC, and also the modifications in the femtocells in the local network. Since their solution is a two-level routing approach. The highest level is carried out by the Mobile Network Layer (MNL), while the lowest-level routing is carried out by the Transport Network Layer (TNL).

Also in [9], the authors started their paper by introducing the LFGW functionalities that provide an architectural connection between the Femtocells. In the second part, the authors explained the concept of NoFs and their functionalities as opposed to the stand alone Femtocells. Then, the authors proposed their architectural solution for the integration of NoFs in the 3GPP Evolved Packet System (EPS). They proposed it by an overview of the NoFs system architecture and its supporting functional entities such as the LFGW and the Modified HeNBs.

Also in [9], the authors presented the main traffic and mobility management challenges that can arise in a NoF, these challenges could be classified into two main problems according to whether they belong to the MNL or TNL. Their solution's improvements to the Mobile Network Layer (MNL) was in determining the GPRS Tunnelling Protocol (GTP) endpoints by their newly designed local location and hand off management mechanisms in order to provide user location and session continuity whilst reducing the volume of signalling traffic that reaches the functional entities in the EPC. Their solution's improvements to the Transport Network Layer (TNL) was in finding the routing path towards the endpoint by the use of Back-pressure based distributed routing mechanism.

## 5. CONCLUSION

This survey paper has provided a detailed explanation of the concepts of QoS in LTE, heterogeneous LTE-A networks, and how traffic is classified within a packet scheduling architecture for heterogeneous LTE-A networks. Then, it summarised the proposed QoS-based packet scheduling algorithms in the literature for Heterogeneous LTE-A Networks, and for Femtocells LTE-A Networks. The importance of this paper, is laying the foundation of understanding the technologies which will be further studied and modelled as a part of the 5G network and beyond 5G.

### CONFLICTS OF INTEREST

The authors declare no conflict of interest.

### APPENDIX A: LTE PROTOCOL STACKS

LTE's radio protocol architecture is separated into two protocol stacks, the first one is the control plane protocol stacks and the second one is the user plane protocol stack as shown in Figure A1 and Figure A2. In these two figures, only the protocols with white background were designed by 3GPP, the others were designed by IETF [10]. In addition, Figure A3 displays the LTE protocol architecture between the eNodeB and the UE with the functionalities of each sub-layers. An example of LTE data flow for three IP packets in the downlink is displayed in Figure A4 "the case of uplink is similar" [7].

The Control plane protocol stack includes the Radio Resource Control layer (RRC) which handles radio-specific functionality that depends on the UE's two modes, either idle or connected. In the case of idle mode, the UE keeps on monitoring the paging channel for detecting incoming calls and acquiring system information. In this mode, control plane protocols include cell selection and re-selection procedures. In the connected mode downlink channel quality and neighbor cell information are being transmitted by the UE to the E-UTRAN to aid it to select the most suitable cell for the UE. In this mode, control plane protocol includes the RRC protocol [22]. The RRC protocol manages UE's signaling and data connections, and it also includes functions for handover [10].

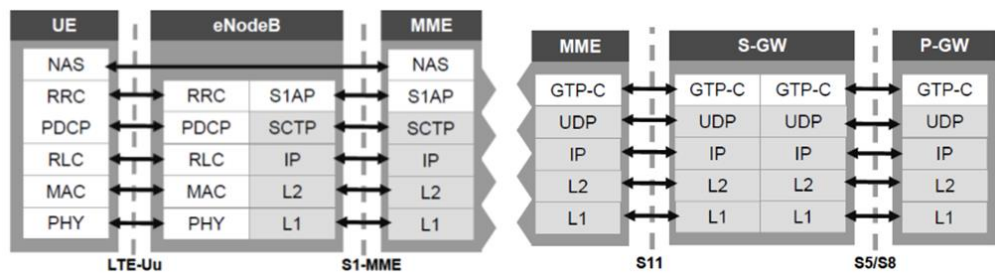

Figure A1. Control plane protocol stack [10]

The topmost layer in the control plane protocol stack is the Non-Access Stratum (NAS), this protocol's transactions contents is only visible to the UE and the MME, it consists of two separate protocols, the EPS Mobility Management (EMM) and the EPS Session Management (ESM). The responsibility of the EPS Mobility Management (EMM) protocol is to handle the UE mobility within the system. When the UE is in idle mode, the EMM protocol attaches the UE to the network and detach it from the network, it also keeps track of the UE location in a process called

Tracking Area Updating (TAU). It also includes the functionalities of re-activating the UE from its idle mode which has two scenarios that differ from the initiating party, one scenario is UE initiated based which is called service request, the other is network initiated based which is called paging. The EMM protocol also authenticates and protect the UE identity, controls the NAS layer security functions, encrypt and protect the integrity of users. The EPS Session Management (ESM) protocol is used for E-UTRAN bearer management procedures in the case if the bearer contexts are not available in the network and E-UTRAN procedures can not start immediately [10].

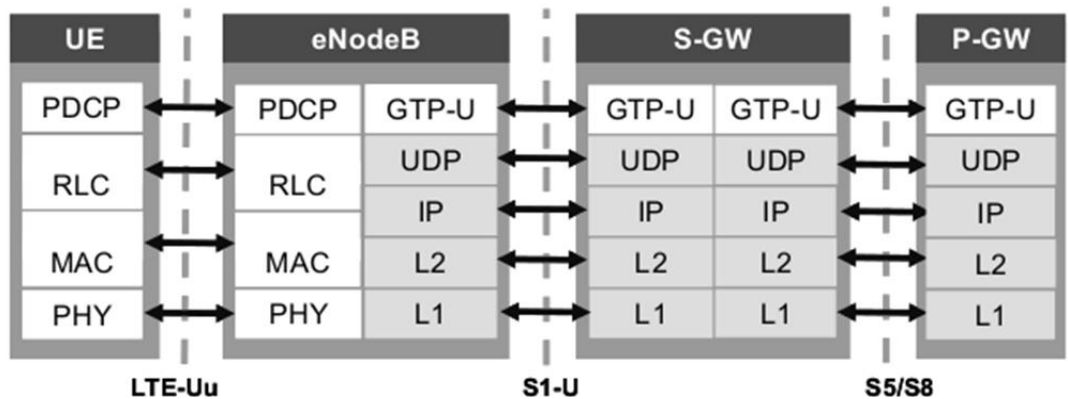

Figure A2. User plane protocol stack [10]

The User plane protocol stack between the UE and eNodeB consists of the following sub-layers [7][10]:

*Packet Data Convergence Protocol (PDCP)*, this protocol's responsibility varies according to the plane type. In the user plane, it is responsible of IP header compression to reduce the amount of bits to be transmitted over the air interface based on a compression technique known as Robust Header Compression (ROHC). In the control plane, it performs encryption, ciphering and deciphering, integrity protection, and also it is responsible of in-sequence delivery and duplicate removal for handover.

*Radio Link Control (RLC)*, this protocol segments and concatenates the PDCP-Protocol Data Units (PDCP-PDUs) for radio interface transmission. It performs error correction with the Automatic Repeat Request (ARQ) method. In addition, it provides services to the RLC in the form of radio bearers. For each UE there is one configured RLC entity per radio bearer.

*Medium Access Control (MAC)*, this layer performs scheduling which is hosted in the eNodeB for both downlink and uplink, multiplex the data into Layer 1 transport blocks, performs error correction with Hybrid ARQ, and it provides services to the RLC in the form of logical channels.

*PHYsical layer (PHY)*, it is the Layer 1 of LTE-Uu radio interface, it performs the typical physical layer functions such as, coding and decoding, modulation and demodulation, and multi-antenna mapping. In addition, it provides services to the MAC layer in the form of transport channels.

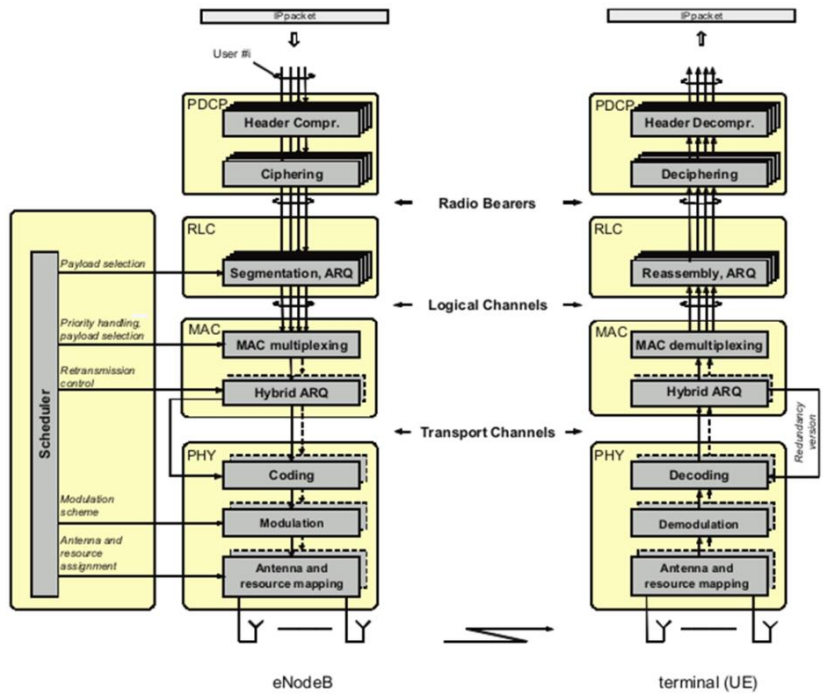

Figure A3. LTE protocol architecture between the eNodeB and UE (downlink) [7]

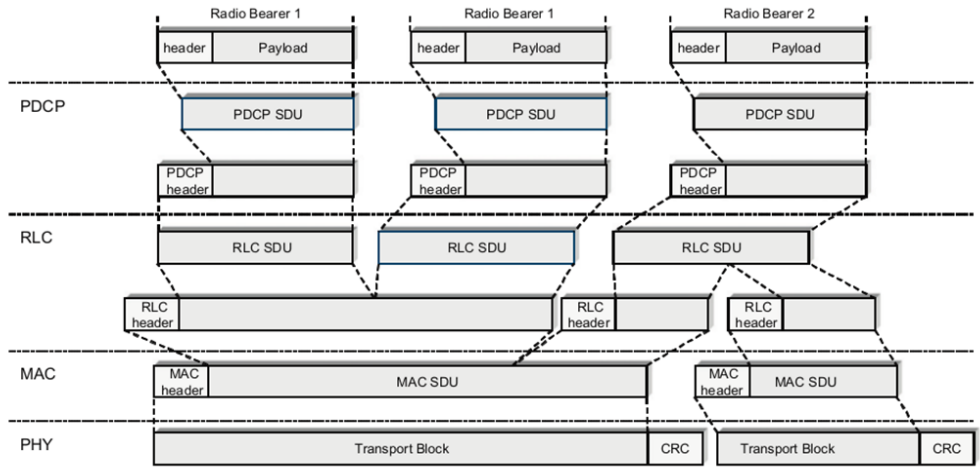

Figure A4. Example of LTE data flow [7]